\documentclass{article}
\usepackage{amssymb,amsmath}
\usepackage{graphicx,enumerate}

\title{How the investor's  risk preferences influence the optimal allocation in a credibilistic portfolio problem}

\author{Irina Georgescu \\ \footnotesize Academy of Economic Studies\\ \footnotesize Department of Economic Cybernetics\\ \footnotesize Pia$\c{t}$a Romana No 6  R 70167, Oficiul Postal 22, Bucharest, Romania\\
 \footnotesize Email: irina.georgescu@csie.ase.ro \\ and \\ Jani Kinnunen \\  \footnotesize
Abo Akademi University, Tuomiokirkkotori 3,
Turku, 20500, Finland\\
\footnotesize Email: jani.kinnunen@abo.fi}

\date{}

\begin{document}
\maketitle

\begin{abstract}

A classical portfolio theory deals with finding the optimal proportion in which an agent invests a
wealth in a risk-free asset and a probabilistic risky asset. Formulating and solving the problem depend on how the risk is represented and how, combined with the utility function defines a notion of expected utility.

In this paper the risk is a fuzzy variable and the notion of expected utility is defined in the setting of Liu's credibility theory. Thus the portfolio choice problem is formulated as an optimization problem in which the objective function is a credibilistic expected utility. Different approximation calculation formulas for the optimal allocation of the credibilistic risky asset are proved. These formulas contain two types of parameters: various credibilistic moments associated with fuzzy variables (expected value, variance, skewness and kurtosis) and the risk aversion, prudence and temperance indicators of the utility function.

\end{abstract}


\newtheorem{definitie}{Definition}[section]
\newtheorem{propozitie}[definitie]{Proposition}
\newtheorem{remarca}[definitie]{Remark}
\newtheorem{exemplu}[definitie]{Example}
\newtheorem{intrebare}[definitie]{Open question}
\newtheorem{lema}[definitie]{Lemma}
\newtheorem{teorema}[definitie]{Theorem}
\newtheorem{corolar}[definitie]{Corollary}

\newenvironment{proof}{\noindent\textbf{Proof.}}{\hfill\rule{2mm}{2mm}\vspace*{5mm}}



%

\section{Introduction}
In classical portfolio theory \cite{eeckhoudt2}, \cite{gollier} it is studied the question of finding the optimum proportion in which an agent invests his wealth
in a risk-free asset and a risky asset. The mathematical model is an optimization problem, formulated in the classical expected utility theory. For the solution of the optimization problem different approximate calculation formulas have been proposed. These formulas use Taylor-type approximations and depend on the investor's risk preferences, as well as on moments of random variables which describe the return of the risky asset.

In \cite{eeckhoudt2}, Chapter 4 and \cite{gollier}, Chapter 5 there are obtained solutions which depend on the expected value, variance and the Arrow-Pratt index \cite{arrow}, \cite{pratt} of the investor's utility function. Papers \cite{athayde}, \cite{garlappi} propose solutions in which the Arrow-Pratt index $r_u$, the prudence index $P_u$ \cite{kimball} and the first three moments appear. A form of the solution depending on the first four moments and the indices of risk aversion, prudence and temperance has been proved in \cite{niguez}. Finally, in \cite{courtois} we find another form of the solution depending on the first four moments.

All portfolio models remembered above are probabilistic: the return of the risky asset is represented by a random variable, and total utility functions are defined as probabilistic expected utilities \cite{eeckhoudt2}, \cite{gollier}. Zadeh's possibility theory \cite{zadeh} offers another way to model portfolio problems: the risky asset is a possibility distribution (more often a fuzzy number), and the formulation and the study of maximization
problems are done in the framework of a possibilistic $EU$-theory (see \cite{carlsson}, \cite{dubois}, \cite{georgescu1}).

In particular, paper \cite{georgescu2} deals with a few portfolio choice models defined in the possibilistic framework: the return of the risky asset is a fuzzy number \cite{dubois}. For these models approximate calculation formulas are found depending on the indicators $r_u$ and $P_u$ as well as on the possibilistic indicators associated with a fuzzy number \cite{carlsson}, \cite{dubois}, \cite{georgescu1}.

Liu's credibility theory developed in \cite{liu1}, \cite{liu2} proved to be an appropriate framework to study the portfolio problem (see monograph \cite{huang}). Using the notion of credibility measure instead of probability or possibility one can develop a credibilistic $EU$-theory, in whose center lies the concept of credibilistic expected utility.

This paper deals with a portfolio problem formulated in a credibilistic context: the return of the risky asset is a fuzzy variable, and the total utility is expressed by the notion of credibilistic expected utility \cite{georgescu1}, \cite{georgescu3}.

Different approximate calculation formulas for the optimal allocation of the credibilistic risky asset depending on the credibilistic moments and risk aversion, prudence and temperance are proved.

The credibilistic moments are obtained from the credibilistic expected utility formula for some particular form of the utility function. The indicators of risk aversion, prudence and temperance reflect some attitudes of the agent faced with risk \cite{eeckhoudt2}, \cite{gollier}, \cite{kimball}, \cite{kimball1}.

We will present shortly the content of the paper.

In Section $2$ there are recalled from \cite{liu2}, \cite{liu1}, \cite{georgescu1} a few basic notions from credibility theory: credibility measure,
credibility space, membership function, credibility distribution, etc. Following \cite{georgescu1} one introduces the credibilistic expected utility, the fundamental notion in the formulation and solving of the credibilistic portfolio problem. By particularization, the formulas of credibilistic expected value, credibilistic variance and other credibilistic moments are obtained.

In Section $3$ the credibilistic standard portfolio problem is defined, analogous to the standard probabilistic problem \cite{eeckhoudt2}, \cite{gollier} and the standard possibilistic problem \cite{georgescu2}. The form of the optimization problem is written in case of a small risk, then a general formula
of the first order condition associated with the optimization problem is obtained.

Section $4$ deals with finding an approximate calculation formula of the solution of the optimization problem using a second order Taylor approximation. In the expression of the approximate solution the first three credibilistic moments and the indicators of absolute risk aversion and prudence appear.

In Section $5$ one shows how one can compute the approximate solution of the optimization problem such that in its expression appear the first four moments and the indicators of absolute risk aversion, prudence and temperance. The main result of the section is an approximation formula for the optimal allocation, where, besides risk aversion and prudence, there is also the investor's temperance.

\section{Credibilistic Expected Utility}

Let $\bf{R}$ be the set of the real numbers, $\Omega \subseteq \bf{R}$  and $\mathcal{P}(\Omega)$ the power set of $\Omega$. The elements of $\Omega$ are called states, and the elements of $\mathcal{P}(\Omega)$ are called events. $A^C$ is the complement of the event $A$.

By \cite{liu1}, a \emph{credibility measure} on $\Omega$  is a function $Cr: \mathcal{P}(\Omega) \rightarrow [0, 1]$ satisfying the following axioms:

($Cr1$) $Cr(\Omega)=1$;

($Cr2$) For any events $A, B$, if $A \subseteq B$ then $Cr(A) \leq Cr(B)$;

($Cr3$) For any event $A$, $Cr(A)+Cr(A^C)=1$;

($Cr4$) For any family of events $(A_i)_{i \in I}$ with the property $\displaystyle \sup_{i \in I} Cr(A_i) <\frac{1}{2}$ the identity $\displaystyle Cr(\bigcup_{i \in I} A_i)=\sup_{i \in I} Cr(A_i)$ holds.

For any event $A$, $Cr(A)$ will be called credibility of $A$.

The triple $(\Omega, \mathcal{P}(\Omega), Cr)$ is called {\emph{credibility space}}. Any probability measure on $\Omega$ is a credibility measure and any probability space is a credibility space. Credibility theory is in a relationship of parallelism with probability theory, but most of the time the definition of the notions and the proofs of the results are very different (see \cite{liu1}). An arbitrary function $\xi: \Omega \rightarrow {\bf{R}}$ will be called {\emph{fuzzy variable}}. The function $\mu : {\bf{R}} \rightarrow [0, 1]$ defined by $\mu(x)=\min(2Cr(\xi=x),1)$, for any $x \in {\bf{R}}$, is
{\emph{the membership function}} associated with $\xi$.

We fix a fuzzy variable $\xi: \Omega \rightarrow {\bf{R}}$ with the membership function $\mu : {\bf{R}} \rightarrow [0, 1]$. The {\emph{credibility distribution}} $\Phi: {\bf{R}} \rightarrow [0, 1]$ of $\xi$ is defined by

$\Phi(x)=Cr(\xi \leq x)$ for any $x \in {\bf{R}}$. (2.1)

Let $u: {\bf{R}} \rightarrow {\bf{R}}$ be a utility function. The {\emph{credibilistic expected utility}} $Q(u(\xi))$ is defined by

$Q(u(\xi))=\int_0^\infty Cr(u(\xi) \geq t)dt-\int_{-\infty}^0 Cr(u(\xi) \leq t)dt$ (2.2)

provided that the two integrals are finite.

We will assume throughout the paper that the following conditions are satisfied:

$\displaystyle \lim_{x \rightarrow -\infty} \Phi(x)=0$; $\displaystyle \lim_{x \rightarrow \infty} \Phi(x)=1$ (2.3)

Then for each monotone utility function $u$ the credibilistic expected utility $Q(u(\xi))$ is written as Stieltjes integral

$Q(u(\xi))=\int_{-\infty}^\infty  u(x)d\Phi(x)$ (2.4)

When $u$ is ${\bf{R}}$'s identity function, one obtains from (2.2) the notion of credibilistic expected value:

$Q(\xi)=\int_0^\infty Cr(\xi \geq t)dt -\int_{-\infty}^0 Cr(\xi \leq t)dt$ (2.5)

For any integer $k \geq 1$ we define the following credibilistic indicators

$\bullet$ $Q(\xi^k)$: the $k$-th order moment of $\xi$;

$\bullet$ $Q((\xi-Q(\xi))^k)$: the $k$-th order central moment of $\xi$.

$V(\xi)=Q((\xi-Q(\xi))^2)$ will be called the credibilistic variance of $\xi$.

A function $f: {\bf{R}} \rightarrow [0, 1]$ with the property $\Phi(x)=\int_{-\infty}^x f(t)dt$ for any $x \in {\bf{R}}$ is called the {\emph{credibility density function}} of $\xi$. In case of the existence of such function, (2.4) gets the form

$Q(u(\xi))=\int_{-\infty}^\infty u(x) f(x) dx$ (2.6)

\begin{propozitie}
Let $g :{\bf{R}} \rightarrow {\bf{R}}$,  $h :{\bf{R}} \rightarrow {\bf{R}}$ be two utility functions and $a, b \in {\bf{R}}$. By denoting $u=ag+bh$ we will have $Q(u(\xi))=aQ(g(\xi))+bQ(h(\xi))$.
\end{propozitie}

The previous proposition expresses the linearity of the credibilistic expected value operator. This property will be used in the next sections without being mentioned.

\section{Credibilistic Standard Model}

One considers an agent who invests a wealth $w_0$ in a risk-free asset and in a risky asset. The agent invests the amount $\alpha$ in the risky asset (stocks) and $w_0-\alpha$ in the risk-free asset (bonds). Let $r$ be the return of the risk-free asset and $x$ the return of the risky asset. According to \cite{eeckhoudt2}, p. 55-56, the value of the portfolio $(w_0-\alpha, \alpha)$ will be given by:

$(w_0-\alpha)(1+r)+\alpha(1+x)=w+\alpha(x-r)$ (3.1)

We denote by $w=w_0(1+r)$ the future wealth of the risk-free strategy.

The agent will have a utility function $u$ of class $\mathcal{C}^2$, increasing and concave. Starting from (3.1), we can formulate various portfolio choice problems depending on the mathematical modeling of the return of the risky asset and the way we define the notion of expected utility.

A probabilistic investment model presented in \cite{eeckhoudt2}, Chapter $4$ or \cite{gollier}, Chapter $5$ is based on the hypothesis that the return of the risky asset is  random variable $X_0$. If we are situated in the framework of von Neumann-Morgenstern expected utility theory, then (3.1) leads to the following maximization problem:

$\displaystyle \max_{\alpha} E[u(w+\alpha(X_0-r))]$ (3.2)

Denoting by $X=X_0-r$ the excess return, (3.2) is written:

$\displaystyle \max_{\alpha} E[u(w+\alpha X)]$ (3.3)

Starting from (3.1), we intend to define a credibilistic portfolio problem. We will assume that the return of the risky asset is a fuzzy variable $\zeta_0$. Using the credibilistic expected utility (2.2), the following maximization problem will be obtained:

$\displaystyle \max_{\alpha} Q[u(w+\alpha(\zeta_0-r))]$ (3.4)

Denoting by $\zeta=\zeta_0-r$ the (credibilistic) excess return, (3.4) will be written:

$\displaystyle \max_{\alpha} Q[u(w+\alpha \zeta)]$ (3.5)

The total utility function in (3.5) is $V(\alpha)=Q[u(w+\alpha \zeta)]$. We will assume the conditions (2.3) fulfilled, thus

$V(\alpha)=\int_{-\infty}^\infty u(w+\alpha x)d\Phi(x)$ (3.6)

Deriving twice, it follows $V''(\alpha)=\int_{-\infty}^\infty x^2 u''(w+\alpha x)d\Phi (x)$. Since $u''\leq 0$, we will have $V''(\alpha) \leq 0$, thus $V$ is concave.

We will study problem (3.5) in case of a small credibilistic risk, assuming that $\zeta=k \mu +\xi$, where $\mu >0$ and $\xi$ is a fuzzy variable such that $Q(\xi)=0$\footnote{For the treatment of portfolio choice problem with probabilistic small risk, see \cite{eeckhoudt2}, \cite{gollier}, \cite{niguez} etc. A study of possibilistic models with small risk can be found in paper \cite{georgescu2}.}. Then

$V(\alpha)=Q[u(w+\alpha(k \mu+\xi)]$  (3.7)

Under conditions (2.3) we have:

$V(\alpha)=\int_{-\infty}^\infty  u(w+\alpha (k \mu+x))d\Phi(x)$ (3.8)

from where it follows immediately

$V'(\alpha)=\int_{-\infty}^\infty (k \mu+x)u'(w+\alpha(k \mu+x))d \Phi (x)$, thus

$V'(\alpha)=Q[(k \mu+\xi)u'(w+\alpha(k \mu+\xi))]$ (3.9)

Let $\alpha(k)$ be the solution of the problem $\displaystyle \max_{\alpha} V(\alpha)$, with $V(\alpha)$ written in form (3.8). Then the first order condition $V'(\alpha(k))=0$ can be written:

$Q[(k \mu+\xi)u'(w+\alpha(k \mu+\xi))]=0$ (3.10)

Similar with the probabilistic case (\cite{gollier}, Section 5.2), we will assume that $\alpha(0)=0$.

The solution of the equation (3.10) is the optimal value of the amount the agent will have to invest in the risky asset. As the exact solution of (3.10) is difficult to determine, we will look for formulas to approximate it.

We will look for approximations of the solution $\alpha(k)$ by formulas such as:

$\alpha(k) \approx \displaystyle \sum_{j=0}^n \frac{k^j}{j!} \alpha^{(j)}(0)$ (3.11)

We consider the Taylor approximation of the derivative $u'(w+\alpha(k \mu+x))$.

$u'(w+\alpha(k \mu+x)) \approx \displaystyle \sum_{j=0}^n \frac{u^{(j+1)}(w)}{j!}\alpha^j (k\mu+x)^j$  (3.12)

We multiply both members of (3.12) by $k \mu+x$:

$(k \mu+x) u'(w+\alpha(k \mu+x)) \approx \displaystyle \sum_{j=0}^n \frac{u^{(j+1)}(w)}{j!}\alpha^j (k\mu+x)^{j+1}$ (3.13)

Taking into account (3.13) and the linearity of credibilistic expected utility operator, we will have:

$Q[(k \mu+\xi)u'(w+\alpha(k \mu+\xi))] \approx \displaystyle \sum_{j=0}^n \frac{u^{(j+1)}(w)}{j!}\alpha^j Q[(k\mu+\xi)^{j+1}]=0$

Then the first order condition (3.10) is written:

$ \displaystyle \sum_{j=0}^n \frac{u^{(j+1)}(w)}{j!}(\alpha(k))^j Q[(k\mu+\xi)^{j+1}]=0$  (3.14)

Using the equation (3.14) we are going to find approximate calculation formulas for the coefficients $\alpha^{(j)}, j=0, \ldots, n$ of (3.11).

For each natural number $n \geq 1$, we will obtain from (3.14) an approximation of $\alpha(k)$.

\section{The Effect of Absolute Risk Aversion and Prudence on the Optimal Allocation}

In this section we will establish an approximate calculation formula for the solution $\alpha(k)$ of equation (3.10), that depends on the absolute risk aversion and prudence indicators associated with the utility function $u$. We recall from that the absolute risk aversion and the prudence of the agent are expressed by the following indicators:

$r_u(w)=-\frac{u''(w)}{u'(w)}$ (the Arrow-Pratt index \cite{arrow}, \cite{pratt}) (4.1)

$P_u(w)=-\frac{u'''(w)}{u''(w)}$ (the Kimball prudence index \cite{kimball}) (4.2)

We will consider the second order Taylor approximation of $\alpha(k)$ around $k=0$:

$\alpha(k) \approx \alpha(0)+k \alpha'(0)+\frac{1}{2} k^2 \alpha''(0)=k \alpha'(0)+\frac{1}{2} k^2 \alpha''(0)$  (4.3)

and we will determine the approximate values of $\alpha'(0)$ and $\alpha''(0)$ using equation (3.14).

\begin{propozitie}
$\alpha'(0) \approx \frac{\mu}{Q(\xi^2)} \frac{1}{r_u(w)}$.
\end{propozitie}

\begin{proof}
For $n=1$, equation (3.14) is written

$u'(w)(k \mu+Q(\xi))+\alpha(k) u''(w)Q[(k \mu+\xi)^2]\approx 0$

Deriving this equation

$u'(w)\mu+u''(w)[\alpha'(k)Q[(k \mu+\xi)^2]+2\alpha(k)\mu Q(k \mu+\xi)] \approx 0$

and setting $k=0$, $\alpha'(0)$ is determined:

$\alpha'(0) \approx -\frac{\mu}{Q(\xi^2)}\frac{u'(w)}{u''(w)}=\frac{\mu}{Q(\xi^2)} \frac{1}{r_u(w)}$.
\end{proof}

\begin{remarca}
In case $n=2$ the approximate solution $\alpha(k)$ will be

$\alpha(k) \approx k \alpha'(0)=\frac{\mu k}{Q(\xi^2)}\frac{1}{r_u(w)}$

depending only on the Arrow-Pratt index.
\end{remarca}

To see how the prudence index $P_u(w)$ appears in the solution $\alpha(k)$ we will have to compute $\alpha''(0)$.

\begin{propozitie}
$\alpha''(0) \approx \frac{P_u(w)}{(r_u(w))^2} \frac{Q(\xi^3)}{(Q(\xi^2))^3} \mu^2$.
\end{propozitie}

\begin{proof}
We write equation (3.14) for $n=2$:

$u'(w)(k \mu+Q(\xi))+u''(w)\alpha(k)Q[(k \mu+\xi)^2]+\frac{u'''(w)}{2}(\alpha(k))^2Q[(k \mu+\xi)^2] \approx 0$

Deriving twice this equation, setting $k=0$ and taking into account that $\alpha(0)=0$, $Q(\xi)=0$, it will follow:

$u''(w)\alpha''(0)Q(\xi^2)+u'''(w)(\alpha'(0))^2Q(\xi^2)\approx 0$

From this equation we obtain for $\alpha''(0)$ the following expression:

$\alpha''(0) \approx -\frac{u'''(w)}{u''(w)}\frac{Q(\xi^3)}{Q(\xi^2)}(\alpha'(0))^2$

Replacing $\alpha'(0)$ with the value from Proposition 4.1 one obtains

$\alpha'(0) \approx \frac{P_u(w)}{(r_u(w))^2}\frac{Q(\xi^3)}{(Q(\xi^2))^3} \mu^2$.
\end{proof}

For $n=3$ we will be able to write the formula for the approximate value of $\alpha(k)$:

\begin{teorema}
$\alpha(k) \approx \frac{k\mu}{Q(\xi^2)}\frac{1}{r_u(w)}+\frac{1}{2}(k \mu)^2 \frac{P_u(w)}{(r_u(w))^2}\frac{Q(\xi^3)}{(Q(\xi^2))^3}$
\end{teorema}

\begin{proof}
$\alpha'(0)$ and $\alpha''(0)$ are replaced in (4.3) with their approximate values given by Propositions 4.1 and 4.3.
\end{proof}

\begin{remarca}
We recall from the previous section that $\zeta=k \mu+\xi$ and $Q(\xi)=0$, thus $Q(\zeta)=k \mu$. Also we notice that $Q(\xi^2)=Q[(\zeta-Q(\zeta))^2]=V(\zeta)$.
\end{remarca}

With these remarks, the solution $\alpha(k)$ from Proposition 4.4 will get the following form:

$\alpha(k) \approx \frac{1}{r_u(w)}\frac{Q(\zeta)}{V(\zeta)}+\frac{1}{2}\frac{P_u(w)}{(r_u(w))^2}\frac{Q[(\zeta-Q(\zeta))^3]}{(V(\zeta))^3}(Q(\zeta))^2$ (4.4)

According to \cite{bhat}, Definition 2.17 or \cite{fono}, Definition 3, the {\emph{credibilistic skewness}} $Sk(\zeta)$ of the fuzzy variable $\zeta$ is defined by

$Sk(\zeta)=Q[(\zeta-Q(\zeta))^3]$ (4.5)

provided that $Q(\zeta)< \infty$. The definition of credibilistic skewness from \cite{ravin} is equivalent to the above definition of $Sk(\zeta)$ (see \cite{ravin}, Theorem III.6).

\begin{remarca}
In other papers (\cite{thav1}, \cite{thav2}) other definitions are introduced for the skewness of a fuzzy variable $\zeta$. They are not equivalent to the one from (4.5), but they are more suitable in the context of this paper.
\end{remarca}

Taking into account (4.5), formula (4.4) can be written as

$\alpha(k) \approx \frac{1}{r_u(w)} \frac{Q(\zeta)}{V(\zeta)}+\frac{1}{2}\frac{P_u(w)}{(r_u(w))^2} \frac{Sk(\zeta)}{(V(\zeta))^3}(Q(\zeta))^2$. (4.6)

In the expression of $\alpha(k)$ we find the indicators of absolute risk aversion and prudence as well as the credibilistic indicators associated with
the fuzzy variable $\zeta$ (expected value, variance and skewness).

\begin{exemplu}
We will assume that the utility function $u$ is HARA-type (\cite{gollier}, Section 3.6).

$u(w)=\theta (\eta+\frac{w}{\delta})^{1-\gamma}$ for $\eta+\frac{w}{\delta}>0$ (4.7)

By \cite{gollier} Section 3.6, the indices of possibilistic risk aversion and prudence associated with this utility function are:

$r_u(w)=(\eta+\frac{w}{\delta})^{-1}$; $P_u(w)=\frac{\delta+1}{\delta}(\eta+\frac{w}{\delta})^{-1}$  (4.8)

from which it follows

$\frac{1}{r_u(w)}=\eta+\frac{w}{\delta}$; $\frac{P_u(w)}{(r_u(w))^2}=\frac{\delta+1}{\delta}(\eta+\frac{w}{\delta})$ (4.9)

Taking into account (4.9), formula (4.6) is particularized as follows:

$\alpha(k) \approx (\eta+\frac{w}{\gamma}) \frac{Q(\zeta)}{V(\zeta)}+\frac{1}{2} \frac{\gamma+1}{\gamma}(\eta+\frac{w}{\gamma})\frac{Sk(\zeta)}{(V(\zeta))^3}{(Q(\zeta))^2}$ (4.10)

Let us consider that $\zeta=(a, b, c)$ is a triangular fuzzy variable (\cite{liu1}) with $a<b<c$. We denote $\alpha=\max(b-a, c-b)$ and $\delta=\min(b-a, c-b)$. By \cite{bhat}, Theorem 2.2 or \cite{fono}, p. 6, in this case we will have the following expressions of the credibilistic indicators from (4.10):

$Q(\zeta)=\frac{a+2b+c}{4}=e$ (4.11)

$V(\zeta)=\frac{33\alpha^3+21\alpha^2\delta +11 \alpha \delta^2 -\delta^3}{384\alpha}$ (4.12)

$Sk(\zeta)=\frac{(c-a)^2}{32}(c+a+2b)=\frac{(c-a)^2}{8}e$ (4.13)

Replacing $Q(\zeta)$, $V(\zeta)$ and $Sk(\zeta)$ in (4.10) we can find the calculation formula of $\alpha(k)$ when $u$ is HARA-type and $\zeta$ is a triangular fuzzy variable.

From (4.11)-(4.13) it follows:

$\frac{Q(\zeta)}{V(\zeta)}=\frac{384 \alpha e}{33\alpha^3+21\alpha^2\delta +11 \alpha \delta^2 -\delta^3}$

$\frac{Sk(\zeta)}{(V(\zeta))^3}(Q(\zeta))^2=\frac{\frac{(c-a)^2}{8}e}{\frac{(33\alpha^3+21\alpha^2\delta +11 \alpha \delta^2 -\delta^3)^3}{384^3(\alpha)^3}}e^2=$

$=\frac{384^3e^3\alpha^3(c-a)^2}{8(33\alpha^3+21\alpha^2\delta +11 \alpha \delta^2 -\delta^3)^3}$.

Replacing $\frac{Q(\zeta)}{V(\zeta)}$ and $\frac{Sk(\zeta)}{(V(\zeta))^3}(Q(\zeta))^2$ in (4.10) it follows:

$\alpha(k) \approx (\eta+\frac{w}{\gamma})\frac{384 \alpha e}{33\alpha^3+21\alpha^2\delta +11 \alpha \delta^2 -\delta^3}
+\frac{1}{2} \frac{\gamma+1}{\gamma}(\eta+\frac{w}{\gamma})\frac{384^3e^3\alpha^3(c-a)^2}{8(33\alpha^3+21\alpha^2\delta +11 \alpha \delta^2 -\delta^3)^3}$

\end{exemplu}

\section{The Effect of Temperance on the Optimal Allocation}

By \cite{kimball1}, the temperance of an agent is an attitude of moderation in the face of risk. The temperance of an agent with the utility function $u$ is measured by the temperance indicator $T_u$ defined by:

$T_u(w)=-\frac{u^{iv}(w)}{u'''(w)}$ (5.1)

To establish the way the temperance appears in the optimal solution $\alpha(k)$ we will set $n=3$ in (3.11)

$\alpha(k) \approx k \alpha'(0)+\frac{1}{2} k^2 \alpha''(0)+\frac{1}{3!}k^3 \alpha'''(0)$ (5.2)

For $\alpha'(0)$ and $\alpha''(0)$ we found approximate calculation formulas in the previous section. An approximate value of $\alpha'''(0)$ will be computed using the equation from the following proposition.

\begin{propozitie}
$\alpha'''(0) Q(\xi^2)+6 \alpha'(0)\mu^2-3P_u(w)[\alpha'(0)\alpha''(0)Q(\xi^3)+3\mu (\alpha'(0))^2Q(\xi^2)]+\frac{T_u(w)}{P_u(w)}(\alpha'(0))^3Q(\xi^4) \approx 0$
\end{propozitie}

\begin{proof}
We write the first order condition (3.14) for $n=3$:

$u'(w)Q(k \mu+\xi)+u''(w)\alpha(k)Q[(k \mu+\xi)^2]+\frac{u'''(w)}{2!}\alpha^2(k)Q[(k \mu+\xi)^3]+\frac{u^{iv}(w)}{3!}\alpha^3(k)Q[(k \mu+\xi)^4] \approx 0$

If we denote

$T_1(k)=\alpha(k)Q[(k \mu+\xi)^2]$ (5.3)

$T_2(k)=\alpha^2(k)Q[(k \mu+\xi)^3]$ (5.4)

$T_3(k)=\alpha^3(k)Q[(k \mu+\xi)^4]$ (5.5)

then the previous equation becomes

$u'(w)Q(k \mu+\xi)+u''(w)T_1(k)+\frac{u'''(w)}{2!}T_2(k)+\frac{u^{iv}(w)}{3!}T_3(k) \approx 0$ (5.6)

Deriving three times (5.6) with respect to $k$ one obtains

$u''(w)T_1'''(k)+\frac{u'''(w)}{2!}T_2'''(k)+\frac{u^{iv}(w)}{3!}T_3'''(k)=0$ (5.7)

We will set $k=0$ in (5.7):

$u''(w)T_1'''(0)+\frac{u'''(w)}{2!}T_2'''(0)+\frac{u^{iv}(w)}{3!}T_3'''(0)=0$ (5.8)

We recall that if $f, g$ are two real functions then

$(fg)'''=f'''g+3f''g'+3f'g''+fg'''$ (5.9)

Using (5.9) we will compute $T_1'''(0)$, $T_2'''(0)$ and $T_3'''(0)$.

{\emph{The computation of $T_1'''(0)$}}

By (5.9) one has

$T_1'''(k)=\alpha'''(k)Q[(k \mu+\xi)^2]+3 \alpha''(k)\frac{d}{dk}Q[(k \mu+\xi)^2]+3 \alpha'(k)\frac{d^2}{dk^2}Q[(k \mu+\xi)^2]+\alpha(k)\frac{d^3}{dk^3}Q[(k \mu+\xi)^2]$

We notice that

$\frac{d}{dk}Q[(k \mu+\xi)^2]=2 \mu Q(k \mu+\xi)=2 \mu (k \mu +Q(\xi))=2 \mu^2 k$

thus $\frac{d}{dk}Q[(k \mu+\xi)^2] \mid_{k=0} =0$

Also, $\frac{d^2}{dk^2}Q[(k \mu+\xi)^2]=2 \mu^2$. Replacing these values in the expression of $T_1'''(k)$ and taking into account that $\alpha(0)=0$ one obtains

$T_1'''(0)=\alpha'''(0)Q(\xi^2)+6 \alpha'(0)\mu^2$ (5.10)

{\emph{The computation of $T_2'''(0)$}}

We denote $g(k)=\alpha^2(k)$. By (5.9), $T_2'''(k)$ has the following form:

$T_2'''(k)=g'''(k)Q[(k \mu+\xi)^3]+3 g''(k) \frac{d}{dk}Q[(k \mu+\xi)^3]+3 g'(k)\frac{d^2}{dk^2}Q[(k \mu+\xi)^3]+g(k)\frac{d^3}{dk^3}Q[(k \mu+\xi)^3]$

We notice that

$g(k)=\alpha^2(k)$; $g(0)=0$;

$g'(k)=2\alpha'(k)\alpha(k)$; $g'(0)=0$;

$g''(k)=2[\alpha''(k)\alpha(k)+(\alpha'(k))^2]$; $g''(0)=2(\alpha'(0))^2$;

$g'''(k)=2[\alpha'''(k)\alpha(k)+3 \alpha'(k)\alpha''(k)]$; $g'''(0)=6 \alpha'(0)\alpha''(0)$.

Also

$\frac{d}{dk}Q[(k \mu+\xi)^3]=3 \mu Q[(k \mu+\xi)^2]$; $\frac{d^2}{dk^2}Q[(k \mu+\xi)^3]=6 \mu^2 Q(k \mu+\xi)$

from where it follows

$\frac{d}{dk}Q[(k \mu+\xi)^3]\mid_{k=0}=3 \mu Q(\xi^2)$; $\frac{d^2}{dk^2}Q[(k \mu+\xi)^3]\mid_{k=0}=6 \mu^2 Q(\xi)=0$.

Setting $k=0$ in the above expression of $T_2'''(k)$ and considering the previous computations it follows

$T_2'''(0)=g'''(0)Q(\xi^3)+3g''(0)3 \mu Q(\xi^2)=6 \alpha'(0) \alpha''(0)Q(\xi^3)+18 \mu (\alpha'(0))^2 Q(\xi^2)$.

{\emph{The computation of $T_3'''(0)$}}

We denote $h(k)=\alpha^3(k)$. By (5.9), the following form of $T_3'''(k)$ is obtained:

$T_3'''(k)=h'''(k)Q[(k \mu+\xi)^4]+3 h''(k) \frac{d}{dk}Q[(k \mu+\xi)^4] +3 h'(k) \frac{d^2}{dk^2}Q[(k \mu+\xi)^4]+
h(k)\frac{d^3}{dk^3}Q[(k \mu+\xi)^4]$

A simple computation shows that

$h(0)=h'(0)=h''(0)=0$; $h'''(0)=6(\alpha'(0))^3$.

Then

$T_3'''(0)=h'''(0)Q(\xi^4)=6(\alpha'(0))^3Q(\xi^4)$

We replace the found values of $T_1'''(0)$, $T_2'''(0)$ and $T_3'''(0)$ in (5.8):

$u''(w)[\alpha'''(0)Q(\xi^2)+6\alpha'(0)\mu^2]+\frac{u'''(w)}{2}[6 \alpha'(0)\alpha''(0)Q(\xi^3)+18 \mu (\alpha'(0))^2 Q(\xi^2)]+
\frac{u^{iv}(w)}{6}[6(\alpha'(0)^3 Q(\xi^4)]=0$

Dividing by $u''(w)$ and considering that $\frac{u^{iv}(w)}{u''(w)}=\frac{T_u(w)}{P_u(w)}$ it follows the equation from Proposition 5.1.
\end{proof}

Now we want to express the formula from Proposition 5.1 in terms of the credibilistic moments of the initial fuzzy variable $\zeta$. For this we recall that $\zeta=k \mu + \xi$ and $Q(\xi)=0$, thus $Q(\zeta)=k \mu$. From the previous sections we know that $Q(\xi^2)=V(\zeta)$ and $Q(\xi^3)=Sk(\zeta)$.

According to \cite{fono}, Definition 4, the {\emph{credibilistic kurtosis}} $K(\zeta)$ of the fuzzy variable $\zeta$ is defined by $K(\zeta)=Q[(\zeta-Q(\zeta))^4]$, provided that $Q(\zeta) < \infty$. It follows $Q(\xi^4)=K(\zeta)$.

Replacing $Q(\xi^2)$, $Q(\xi^3)$, $Q(\xi^4)$ in the formula from Proposition 5.1 we will obtain

\begin{corolar}
The values $\alpha'(0)$, $\alpha''(0)$ and $\alpha'''(0)$ verify the following dependence relation:

$\alpha'''(0) V(\zeta)+6 \alpha'(0) \mu^2- 3P_u(w)[\alpha'(0)\alpha''(0)Sk(\zeta)+3 \mu (\alpha'(0))^2V(\zeta)]+\frac{T_u(w)}{P_u(w)}(\alpha'(0))^3 K(\zeta) \approx 0$.

The values $\alpha'(0)$, $\alpha''(0)$ are computed with the formulas from Propositions 4.1 and 4.3, which can be written as follows:

$\alpha'(0) \approx \frac{1}{r_u(w)}\frac{\mu}{V(\zeta)}$   (5.11)

$\alpha''(0) \approx \frac{P_u(w)}{(r_u(w))^2}\frac{Sk(\zeta)}{(V(\zeta))^3} \mu^2$ (5.12)

\end{corolar}

Using (5.11) and (5.12), from Corollary 5.2 we can obtain the following result:

\begin{teorema}
The optimal solution $\alpha(k)$ is computed with the formula

$\alpha(k) \approx \frac{1}{r_u(w)} \frac{Q(\zeta)}{V(\zeta)}+\frac{1}{2}\frac{P_u(w)}{(r_u(w))^2}\frac{Sk(\zeta)}{(V(\zeta))^3} ((Q(\zeta))^2-
\frac{1}{r_u(w)} \frac{(Q(\zeta))^3}{(V(\zeta))^2} +\frac{1}{2} \frac{(P_u(w))^2}{(r_u(w))^3} \frac{Q(\zeta)}{(V(\zeta))^5} (Sk(\zeta))^2 +
\frac{3}{2} \frac{P_u(w)}{(r_u(w))^2} \frac{(Q(\zeta))^3}{(V(\zeta))^2} -\frac{1}{3!} \frac{1}{(r_u(w))^3} \frac{T_u(w)}{P_u(w)}
\frac{(Q(\zeta))^3}{(V(\zeta))^4} K(\zeta)$
\end{teorema}

\begin{proof}
We recall that $\alpha(k)$ is computed with the formula:

$\alpha(k) \approx k \alpha'(0)+\frac{1}{2}k^2 \alpha''(0)+\frac{1}{3\, !}\alpha'''(0)$ (5.13)

By (5.11) and (5.12)

$k \alpha'(0)+\frac{1}{2}k^2 \alpha''(0) \approx \frac{1}{r_u(w)} \frac{Q(\zeta)}{V(\zeta)}+\frac{1}{2}\frac{P_u(w)}{(r_u(w))^2}\frac{Sk(\zeta)}{(V(\zeta))^3}(Q(\zeta))^2$ (5.14)

It remains to compute $\frac{1}{3\, !}k^3 \alpha'''(0)$ using (5.11), (5.12) and $Q(\zeta)=k \mu$.

A simple computation shows that

$\bullet$ $\frac{1}{3\, !}k^3 6 \alpha'(0) \mu^2=\frac{1}{r_u(w)} \frac{(Q(\zeta))^3}{V(\zeta)}$

$\bullet$ $\frac{1}{3\, !} k^3 3 P_u(w) \alpha'(0) \alpha''(0)Sk(\zeta)=\frac{1}{2} \frac{(P_u(w))^2}{(r_u(w))^3}\frac{(Q(\zeta))^3}{(V(\zeta))^4}(Sk(\zeta))^2$

$\bullet$ $\frac{1}{3\, !}k^3 9 P_u(w) \mu (\alpha'(0))^2 V(\zeta)=\frac{3}{2} \frac{P_u(w)}{(r_u(w))^2} \frac{(Q(\zeta))^3}{V(\zeta)}$

$\bullet$ $\frac{1}{3\, !}k^3 \frac{T_u(w)}{P_u(w)} (\alpha'(0))^3 K(\zeta)=\frac{1}{3\, !} \frac{1}{(r_u(w))^3} \frac{T_u(w)}{P_u(w)}\frac{(Q(\zeta))^3}{(V(\zeta))^3}K(\zeta)$

Multiplying the identity of Corollary 5.2 by $\frac{1}{3\, !}k^3$ and taking into account the four equalities above, it follows

$\frac{1}{3\, !}k^3 \alpha'''(0) V(\zeta) +\frac{1}{r_u(w)}\frac{((Q(\zeta))^3}{V(\zeta)}-$

$-\frac{1}{2}\frac{(P_u(w))^2}{(r_u(w))^3}\frac{(Q(\zeta))^3}{(V(\zeta))^4}(Sk(\zeta))^2$

$-\frac{3}{2} \frac{P_u(w)}{(r_u(w))^2}\frac{(Q(\zeta))^3}{V(\zeta)}+\frac{1}{3\, !} \frac{1}{(r_u(w))^3}\frac{T_u(w)}{P_u(w)} \frac{(Q(\zeta))^3}{(V(\zeta))^3}K(\zeta) \approx 0$.

From this relation we find the value of $\frac{1}{3\, !}k^3 \alpha'''(0)$:

$\frac{1}{3\, !}k^3 \alpha'''(0) \approx - \frac{1}{r_u(w)}\frac{((Q(\zeta))^3}{(V(\zeta))^2}+\frac{1}{2}\frac{(P_u(w))^2}{(r_u(w))^3}\frac{(Q(\zeta))^3}{(V(\zeta))^5}(Sk(\zeta))^2$

$+\frac{3}{2} \frac{P_u(w)}{(r_u(w))^2}\frac{(Q(\zeta))^3}{(V(\zeta))^2}-\frac{1}{3\, !} \frac{1}{(r_u(w))^3}\frac{T_u(w)}{P_u(w)} \frac{(Q(\zeta))^3}{(V(\zeta))^4}K(\zeta) \approx 0$.

Replacing in (5.13) $k \alpha'(0)+\frac{1}{2}k^2 \alpha''(0)$ with the value from (5.14) and $\frac{1}{3\, !}k^3 \alpha'''(0)$ with the value from the previous identity, it follows the approximate value of $\alpha(k)$ from the enunciation of the theorem.
\end{proof}

\begin{remarca}
In the approximation formula of $\alpha(k)$ from Proposition 4.3 the following parameters appear:

$\bullet$ the credibilistic indicators $Q(\zeta)$, $V(\zeta)$, $Sk(\zeta)$, $K(\zeta)$ associated with the fuzzy variable $\zeta$ (=the credibilistic excess return);

$\bullet$ the investor's risk preference indicators $r_u(w)$, $P_u(w)$ and $T_u(w)$.
\end{remarca}

\begin{exemplu}
Assume that the utility function $u$ is CRRA-type: $u(w)=\frac{w^a}{a}, a>0$. Then

$r_u(w)=\frac{1-a}{w}$, $P_u(w)=\frac{2-a}{w}$, $T_u(w)=\frac{3-a}{w}$

from which it follows:

$\frac{1}{P_u(w)}=\frac{w}{2-a}$, $\frac{P_u(w)}{r_u(w)}=\frac{2-a}{1-a}$, $\frac{P_u(w)}{(r_u(w))^2}=\frac{w(2-a)}{(1-a)^2}$, $\frac{T_u(w)}{P_u(w)}=\frac{3-a}{2-a}$, $\frac{(P_u(w))^2}{(r_u(w))^3}=\frac{w(2-a)^2}{(1-a)^3}$, $\frac{T_u(w)}{P_u(w)((r_u(w))^3}=\frac{w^3(3-a)}{(1-a)^3(2-a)}$.

Replacing in the formula from Theorem 5.3 we find

$\alpha(k) \approx \frac{w}{1-a} \frac{Q(\zeta)}{V(\zeta)}+\frac{1}{2} \frac{w(2-a)}{(1-a)^2}\frac{Sk(\zeta)}{(V(\zeta))^3}
(Q(\zeta))^2-\frac{w}{1-a}\frac{(Q(\zeta))^3}{(V(\zeta))^2}+\frac{1}{2}\frac{(2-a)^2w}{(1-a)^3} \frac{Q(\zeta)}{(V(\zeta))^5} (Sk(\zeta))^2
+\frac{3}{2} \frac{w(2-a)}{(1-a)^2} \frac{(Q(\zeta))3}{(V(\zeta))^2} - \frac{1}{3!} \frac{w^3(3-a)}{(1-a)^3(2-a)} \frac{(Q(\zeta))^3}{(V(\zeta))^4} K(\zeta)$.

We assume that $\zeta$ is a triangular fuzzy variable $(a, b, c)$ with $a<b<c$. Then, keeping the notations from Example 4.7, $Q(\zeta)$, $V(\zeta)$ and $Sk(\zeta)$ are computed with the formulas (4.11), (4.13). By \cite{fono}, Proposition 3 (3), the kurtosis $K(\zeta)$ has the following expression:

$K(\zeta)=\frac{253 \alpha^5+ 395 \alpha^4 \gamma + 17 \alpha \gamma^4 +290 \alpha^3 \gamma^2 + 70 \alpha^2 \gamma^3 - \gamma^5}{10240 \alpha}$ (5.13)

With these expressions of $Q(\zeta)$, $V(\zeta)$, $Sk(\zeta)$ and $K(\zeta)$ and for given values of $w$ and $a$, we will be able to compute the approximate value of $\alpha(k)$.
\end{exemplu}

\section{Final Remarks}

This paper contains the following contributions:

(a) Formulating and analyzing an investment model in the context of a credibilistic expected theory (based on a concept of credibilistic expected utility);

(b) An approximate calculation formula of the optimal allocation in terms of absolute risk aversion and investor's prudence, as well as the following credibilistic indicators: expected value, variance and skewness (Theorem 4.4 or formula (4.6));

(c) a second approximate calculation formula of the optimal allocation, in which besides the above mentioned indicators the investor's temperance and the credibilistic kurtosis appear.

We will mention now two open problems.

(1) In \cite{georgescu2} mixed portfolio problems have been studied: besides the investment risk one considers a background risk. Both the investment risk and the background risk can be a random variable or a fuzzy number. It would be interesting the analysis of such models with background risk, in which the two types of risk include the credibilistic risk. Also it would be realistic to consider models in which the background risk is a multidimensional vector, with probabilistic, possibilistic or credibilistic components.

(2) Let us denote by $\alpha_n(k)$ the approximate solution of the equation (3.14) written as (3.11). Can a recurrent solution be found for the solution $\alpha_n(k)$?



%

\end{document}